\documentclass[twocolumn, prb]{revtex4-1}

\usepackage{graphicx}
\usepackage{url}
\usepackage{textcase}
\usepackage{natbib}
\usepackage{amsmath}
\usepackage{xspace}
\usepackage{longtable}
\usepackage[usenames,dvipsnames]{color}
\usepackage[dvipdfmx]{hyperref}

\graphicspath{{./fig/}}

\newcommand{\Eu}{{Eu$_2$Ir$_2$O$_7$}\xspace}
\newcommand{\Tmn}{T_{\rm{MN}}}

\newcommand{\pyro}[2]{#1$_2$#2$_2$O$_7$\xspace}
\newcommand{\Ea}{E_{\rm{a}}}
\newcommand{\kb}{k_{\rm{B}}}

\begin{document}
\title{Continuous Transition between Antiferromagnetic Insulator and Paramagnetic Metal in the Pyrochlore Iridate Eu$_2$Ir$_2$O$_7$}

\author{Jun J. Ishikawa}
\email{jun@issp.u-tokyo.ac.jp}
\affiliation{Institute for Solid State Physics, University of Tokyo, Kashiwa, Chiba 277-8581, Japan}

\author{Eoin C. T. O'Farrell}
\affiliation{Institute for Solid State Physics, University of Tokyo, Kashiwa, Chiba 277-8581, Japan}

\author{Satoru Nakatsuji}
\email{satoru@issp.u-tokyo.ac.jp}
\affiliation{Institute for Solid State Physics, University of Tokyo, Kashiwa, Chiba 277-8581, Japan}

\date{\today}

\begin{abstract}
Our single crystal study of the magneto-thermal and transport properties of the pyrochlore iridate \Eu reveals 
a continuous phase transition from a paramagnetic metal to an antiferromagnetic insulator for a sample with stoichiometry within $\sim$1 \% resolution. The insulating phase has strong proximity to an antiferromagnetic semimetal, which is stabilized by several \% level of the off-stoichiometry. 
Our observations suggest that in addition to electronic correlation and spin-orbit coupling the magnetic order is essential for opening the charge gap. 
\end{abstract}



\maketitle

The effects of spin-orbit (SO) coupling have been one of the recent central subjects in condensed matter physics.
Along this stream, extensive theoretical and experimental work has been performed on topological insulators and elucidated the gapless helical surface state in weakly interacting semiconductors \cite{Hasan,SCZhang}.
Significant roles of the SO coupling have been also found in various electron correlation effects in the 5$d$ transition metal based oxides.
In particular, the study on the tetragonal iridium oxide Sr$_2$IrO$_4$\cite{kim_prl, kim_science} has indicated that the strong SO coupling splits the otherwise wide $5d~t_{\rm 2g}$ band into two bands based on an effective pseudospin $j_{\rm{eff}}=1/2$ doublet and a $j_{\rm{eff}}=3/2$ quadruplet.
It has been proposed that the half-filled relatively narrow $j_{\rm{eff}}=1/2$ band at the Fermi level ($E_{\rm F}$) opens the Mott gap due to electron correlation.

For the study of the interplay between electron correlation and SO coupling in $5d$ transition metal compounds, pyrochlore oxides are suitable because the corner sharing tetrahedra network tends to form a narrow flat band at $E_{\rm F}$ that enhances the effects of electron correlation and/or SO coupling, both of which have comparable size with the bandwidth. 
In particular, rare-earth pyrochlore iridates \pyro{\textit{R}}{Ir} (\textit{R} = rare-earth elements) provide an ideal class of materials because they have been reported to show a variety of transport and magnetic behaviors \cite{yanagishima,taira,matsu2007,matsu2011}, varying from
a correlated metal\cite{nakatsuji06, machida07, machida_nat} to an insulator\cite{NdIr,tafti} depending on the ionic radius of rare-earth elements. They are also attractive because the one hole nature of Ir$^{4+}$ $5d~ t_{\rm 2g}^5$ state simplifies the electronic structure and allows further in depth comparison between experiment and theory.
Indeed, a number of interesting theoretical proposals have been made as possible ground states of the iridium based pyrochlore oxides, such as a topological insulator, a Weyl semimetal, and an antiferromagnetic insulator
\cite{balents, ybkim2010, wan2011, kurita, krempa}.

To clarify the basic magnetic and transport properties due to the iridium pyrochlore network, 
the simplest to study is the systems with rare-earth elements with nonmagnetic properties, namely, Eu$^{3+}$ or Y$^{3+}$.
While Y$_2$Ir$_2$O$_7$ is known to be insulating at all the temperature measured, the experimental situation for Eu analog is controversial. 
Pioneer work by Yanagishima and Maeno reports that it is a nonmagnetic metal\cite{yanagishima}, while recent work by Matsuhira {\it et al.} found a metal-insulator  transition with a magnetic ordering in the insulating phase\cite{matsu2007, matsu2011}. Recent Raman study found a subtle structural symmetry change\cite{hasegawa} while the single crystal X-ray indicates no sign of the structural transition\cite{millican2007}. This suggests that Eu$_2$Ir$_2$O$_7$ has proximity to a metal-insulator transition and therefore best suited for the study of the interplay between the electronic correlation and spin-orbit coupling. 

In order to reveal the intrinsic behavior of this interesting pyrochlore iridate, we have performed detailed single crystal study of transport and magnetic properties and their systematic dependence on sample quality. It reveals that the ground state of the stoichiometric Eu$_2$Ir$_2$O$_7$ is an antiferromagnetic (AF) insulator with a small charge gap ($\sim 10$ meV) and with magnetic isotropy. The resistivity is exponentially divergent on cooling, excluding the possibility of the topological semi-metallic state predicted by theory  \cite{wan2011, krempa} .
Instead, this insulator has strong proximity to an AF semimetal or metal phase which may be stabilized by carrier doping induced by a few \% off-stoichiometry. The fact that the onset of the AF ordering is independent of the magnitude of the off-stoichiometry indicates that the AF order is essential for opening the charge gap.
These results are fully consistent with the recent theoretical proposal of the antiferromagnetic insulating ground state with ``all-in all-out'' spin structure\cite{wan2011,krempa}.

Single crystals of \Eu were grown by a KF flux method from polycrystalline \Eu,
prepared by solid state reaction of the appropriate mixture of Eu$_2$O$_3$ (4N) and IrO$_2$ (4N) powders \cite{millican2007}.
Powder and single crystal X-ray diffraction analyses confirm single phase and the pyrochlore structure with the lattice constant $a =$ 10.274(3) \AA.
Electron probe microanalysis (EPMA) found a slight deviation up to several \% from the stoichiometry in the ratio between Eu and Ir contents. Our detailed study using synchrotron X-ray measurements of the related rare-earth iridate Pr$_2$Ir$_2$O$_7$ found the excess rare-earth / iridium occupies the site of the counterpart ion without leaving vacancy \cite{Sawa}. Therefore, throughout this paper we adopt the chemical formula Eu$_{2(1-x)}$Ir$_{2(1+x)}$O$_{7+\delta}$. As we will discuss below, our measurements indicate that $\delta$ should be zero, when $x = 0$. The change in the lattice parameter due to the off-stoichiometry was not detected within experimental resolution using our laboratory based X-ray systems. This suggests that the major effect would be the carrier doping due to the valence imbalance between Eu$^{3+}$ and Ir$^{4+}$ ions.

The longitudinal and Hall resistivity was measured by a standard four-probe method from 2 K to 300 K using a variable temperature insert system under a field along [111] up to 9 T.
DC magnetization under a field along [100], [110], and [111] axes was measured using a commercial SQUID magnetometer (MPMS, Quantum Design) from 2 K to 350 K under a magnetic field of 0.1 T in both field cooled (FC) and zero field cooled (ZFC) conditions.
Specific heat was measured by thermal relaxation method using a commercial system (PPMS, Quantum Design).



\begin{figure}[b]
\begin{center}
  \includegraphics[width=\columnwidth]{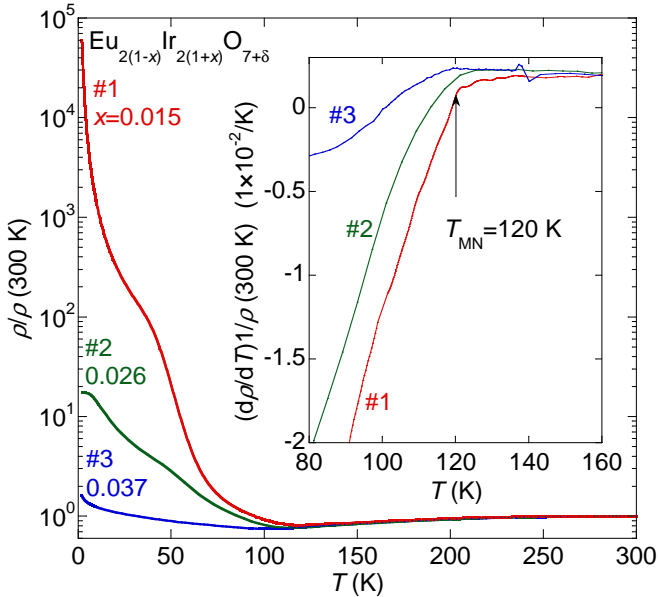}
\caption{(Color online) Temperature dependence of the electrical resistivity of single crystalline \Eu for three samples with different off-stoichiometry $x$, normalized by the value at 300 K. Inset: Temperature dependence of  the temperature derivative of the resistivity $\mathrm{d}\rho/\mathrm{d} T$ normalized by the resistivity at 300 K.} 
\label{fig1_r}
\end{center}
\end{figure}

First we present the temperature dependence of the electrical resistivity $\rho(T)$.
To discuss the strong sensitivity to the off-stoichiometry, 
Figure \ref{fig1_r} shows $\rho(T)$ normalized by its room temperature value, $\rho(T) / \rho(300~{\mathrm K})$, for single crystals with various off-stoichiometry $x$.
In the high temperature region of $T >120$ K, all the data of $\rho(T) / \rho(300~{\mathrm K})$ collapse on top of each other  and shows metallic behavior with positive slope with a typical value of 10 m$\Omega$cm at room temperature.
The Ioffe-Regel limit was estimated to be $\sim 1.4$ m$\Omega$cm, indicating that
transport is strongly incoherent in this high $T$ regime. 

On cooling in the range of $120$ K $>T>100$ K, all the samples commonly show a broad upturn in $\rho(T)$ with an anomaly in $\textrm{d}\rho/\textrm{d} T$ at $T=120$ K (inset of Fig. \ref{fig1_r}).
Here at $T=120$ K we locate the onset of the low temperature nonmetallic (negative $\textrm{d}\rho/\textrm{d} T$) behavior and refer to this as the metal-nonmetal transition temperature $\Tmn$.
No hysteresis in $\rho(T)$ was observed across $\Tmn$, indicating the second-order nature of the transition.

\begin{figure}[t]
 \begin{center}
  \includegraphics[width=\columnwidth]{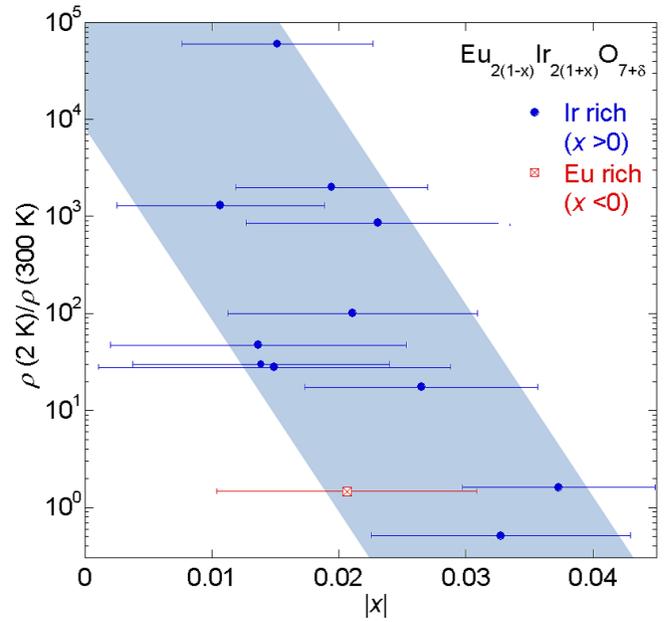}
 \end{center}
 \caption{
(Color online) Inverse resistivity ratio 1/RRR $= \rho(2~{\mathrm K}) / \rho(300~{\mathrm K})$ vs. the off-stoichiometry $\vert x\vert$  for various single crystals of Eu$_{2(1-x)}$Ir$_{2(1+x)}$O$_{7+\delta}$. Blue and red symbols correspond to Ir rich ($x >0$) and Eu rich ($x < 0$)~samples, respectively. The blue shaded belt is
a guide to the eye.
\label{fig4_rrr}}

\end{figure}

Below $\Tmn$, $\rho(T)$ is found substantially different with various residual resistivity ($\rho(T=2$ K)) varying by four orders of magnitude between various samples. To clarify the origin of the sample variation in $\rho(T)$, we plot in Fig. \ref{fig4_rrr} the inverse resistivity ratio 1/RRR $= \rho(2 K) / \rho(300 {\textrm K})$ vs. the off-stoichiometry $x$ for various single crystals.
The deviation from stoichiometry is found at most 5 \% and the majority of samples are Ir rich.
There is a correlation between $\vert x\vert$ and 1/RRR, which suggests that the most stoichiometric samples have the largest 1/RRR \emph{i.e.} being the most insulating.
This also indicates when  $\vert x\vert \rightarrow 0$, $\delta$ should also vanish, so that no doped carriers remain. 
Therefore, we conclude that the intrinsic low temperature state of \Eu is insulating and 1/RRR can be used as figure of merit for the sample quality.

\begin{figure}[t]
 \begin{center}
  \includegraphics[width=\columnwidth]{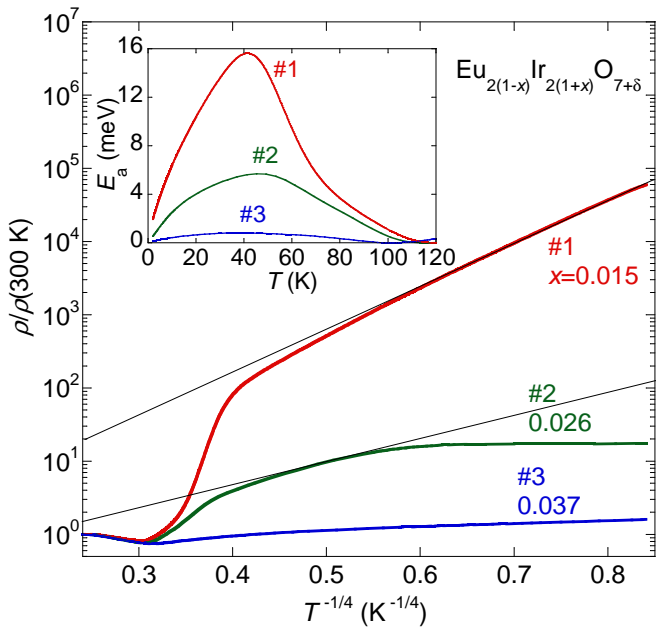}
 \end{center}
\caption{
(Color online) $\ln \rho(T)/\rho(300~{\mathrm K})$ vs. $1/T^{1/4}$ of single crystalline \Eu for three samples with different off-stoichiometry  $x$. The solid lines are guides to the eye. Inset: Temperature dependence of the activation energy $\Ea$.  For definition, see text.
\label{fig_VRH}}
\end{figure}

The insulating behavior can be also confirmed in the temperature dependence of the resistivity $\rho(T)$. Figure \ref{fig1_r} shows that the most stoichiometric crystal \#1 with the smallest value of $\vert x\vert = 0.015(8)$ exhibits the most resistive behavior and a clear shoulder at $\sim 50$ K. As for the crystal \#2 with larger $x = 0.026(9)$, $\rho(T)$ saturates at low temperatures, resulting in a small value of 1/RRR $= \rho(2~{\mathrm K}) / \rho(300~{\rm K})$. The sample with a smaller 1/RRR has a weaker anomaly in $\mathrm{d}\rho/\mathrm{d} T$ at $\Tmn$, thus indicating that off-stoichiometry broadens the transition. For the sample with the largest $x =$ 0.037(8) among three crystals used in Fig. \ref{fig1_r}, $\rho(T)$ shows only a weak kink at $\Tmn$ and display a semi-metallic behavior with a gradual increase of the resistivity on cooling. 
To further characterize the insulating behavior, we plotted $\ln \rho(T)$ vs. $1/T^{\alpha}$ with various $\alpha$. As shown in Fig. \ref{fig_VRH}, the variable range hopping (VRH) $\rho(T)=\rho_0 \exp(T_0/T)^{1/4}$ with $\alpha = 1/4$ for three dimensions is found to best describe the insulating behavior\cite{CaSr214, mottdavis, shklovskii, mott}, in particular, for the crystal  \#1 with $x$=0.015(8) over a decade of $T$ below the shoulder temperature of 50 K and down to the lowest $T$ of the measurement, 2 K. With increasing the off-stoichiometry, $\rho(T)$ deviates from the VRH fit (solid line in Fig. \ref{fig_VRH}) and levels off at low temperatures. This strongly suggests that the insulating state of the near stoichiometric sample with $x$=0.015(8) has strongly localized states at the Fermi level $E_{\rm F}$ induced by the slight off-stoichiometry. A rough estimate of the charge gap can be made using the Arrhenius's law $\rho(T)=\rho_0 \exp(\Ea/\kb T)$ with $\rho_0$ being the minimum of the resistivity. The resultant $T$ dependent $\Ea(T)$ for the crystal \#1, shown in the inset of Fig. \ref{fig_VRH}, reaches its maximum of 16 meV at 40 K, indicating that \Eu is a narrow gap semiconductor.

Comparing our single crystal results to recent results for polycrystalline samples\cite{matsu2011} finds good consistency in the value of 1/RRR $\sim10^5$, providing another evidence for the high quality of our stoichiometric single crystal. More than ten times larger absolute value of the resistivity and the nonmetallic behavior with negative slope found even at $T > \Tmn$ in the polycrystalline samples indicate extrinsic scattering due to the polycrystalline grain boundaries, which is absent in our single crystals.

\begin{figure}[t]
 \begin{center}
  \includegraphics[width=\columnwidth]{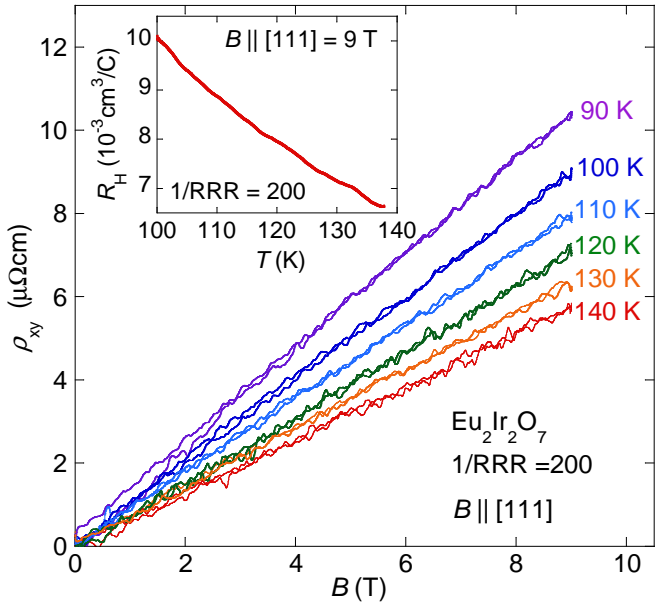}
 \end{center}
\caption{
(Color online) Field dependence of the Hall resistivity $\rho_{xy}$ for single crystalline \Eu with 1/RRR $= 200$ obtained under external field along [111] at various temperatures between 90 K and 140 K across the transition temperature of $\Tmn = 120$ K. Inset: Temperature dependence of the Hall coefficient $R_{\rm H} = \rho_{xy}/B$ obtained under a field of 9 T applied along [111].
\label{fig_Hall}}
\end{figure}

To further reveal the nature of the transition at $\Tmn$, we measured the magnetoresistance and Hall resistivity as a function of field and temperature. 
Except a slight broadening of the kink at $\Tmn$, no major change was observed in $\rho(T)$ under field up to 9 T applied along [111], revealing the robust feature of the transition against the application of the field.
The field dependence of the Hall resistivity $\rho_{xy}$ plotted in Fig. \ref{fig_Hall} shows a linear dependence on magnetic field $\mu_0 H$ up to at least 9 T in the temperature region between 90 and 140 K across $\Tmn$.
This linear response allows us to define the Hall coefficient defined by $R_{\rm H} = \rho_{xy}/\mu_0 H$, whose temperature dependence is plotted in the inset of Fig. \ref{fig_Hall}. Interestingly, no anomaly was found at $\Tmn$ in sharp contrast with the longitudinal resistivity results. This behavior is similar to the one observed for Cd$_2$Os$_2$O$_7$\cite{Mandrus}. As we found above, the charge gap scale is only of the order of 10 meV comparable with the scale of $\Tmn$. Thus, thermal excitations across the gap should be significant at $\sim \Tmn$, and may effectively wipe out the charge gap, smoothening the $T$ dependence of $R_{\rm H}$.

Note that $\rho_{xy}$ at 90 K shows small hysteresis suggestive of time reversal symmetry breaking at 0 T.
Actually, hysteresis in $M$-$H$ curve (not shown) also appears just below $\Tmn$ with evolution of a small spontaneous magnetization.
Hysteresis in $\rho_{xy}$ appears to come from the anomalous component proportional to the magnetization 
due to spin-orbit coupling and/or impurity scattering\cite{Nagaosa, Karplus, Smit1955, Smit1958, Berger}.
Considering the hysteresis in the $MH$ curve together with the one observed for the susceptibility as a function of temperature described below, 
weak spontaneous magnetization may originate from canting of non-collinear Ir moments.

\begin{figure}[t]
 \begin{center}
  \includegraphics[width=\columnwidth]{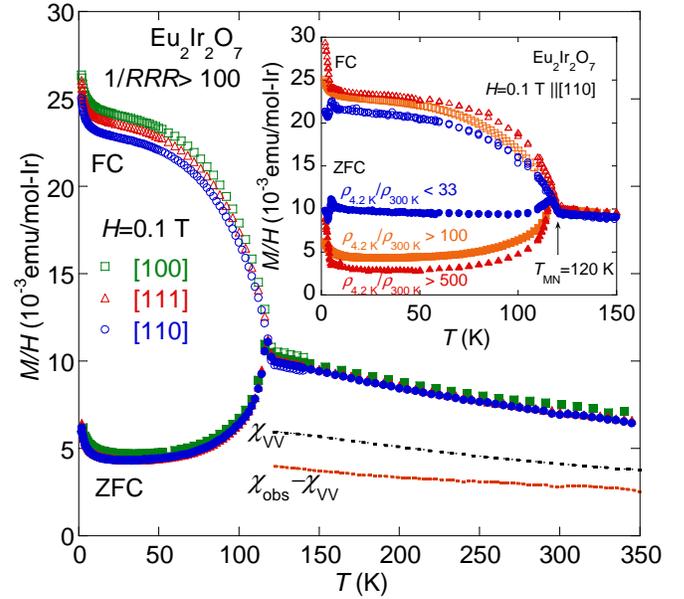}
 \end{center}
 \caption{
(Color online) Temperature dependence of the magnetic susceptibility $\chi = M/H$ of single crystalline \Eu with 1/RRR $> 100$, measured using zero field cool (ZFC) and field cool (FC) sequences under a field of 0.1 T aligned along [100], [111], and [110]. The broken lines indicate the Van-Vleck contribution $\chi_{\textrm{VV}}$ estimated using the excited multiplet levels for a $J=0$ ground state of Eu$^{3+}$ ion and the component obtained after subtracting the Van-Vleck contribution from the observed $\chi(T)$, labeled as $\chi_{\rm obs}- \chi_{\rm VV}$. Inset: $\chi(T)$ for three different qualities of crystals under a field of 0.1 T along [110]. Both FC and ZFC results are shown. \label{fig2_chi}
}

\end{figure}

Figure \ref{fig2_chi} shows the temperature dependence of the magnetic susceptibility $\chi(T) = M(T)/H$ for a batch of 5 single crystals with $\rho(4.2~\mathrm{K})/\rho(300~\mathrm{K}) >500$. The measurement was made using zero field cool (ZFC) and field cool (FC) sequences under a field of 0.1 T aligned along various high symmetry crystal axes.
A sharp transition was observed as a kink in $\chi(T)$ at $T=120$ K, the same temperature as $\Tmn$. In addition, no hysteresis was observed across $\Tmn$, indicating the transition is of the second-order type.
The measurements for other batches of crystals with respective $ \rho(4.2~\mathrm{K})/\rho(300~\mathrm{K}) >100$ and with  $\rho(4.2~\mathrm{K})/\rho(300~\mathrm{K}) < 33$ found the same onset temperature of 120 K, and in particular for the FC sequence, the nearly identical temperature dependence of $\chi(T)$. This is striking given the strong sensitivity of $\rho(T)$ to the off-stoichiometry. The magnetic transition at $\Tmn$ is also in good agreement with the observation of a magnetic order observed by muon spin rotation ($\mu$SR) \cite{zhao}.

Above $\Tmn$, $\chi(T)$ shows a weak temperature dependence. 
Generally for Eu compounds, it is known that the Van-Vleck paramagnetism of Eu$^{3+}$ ions has to be considered because there are relatively low lying $J=1$ ($\sim 480$ K) and $J=2$ ($\sim 1440$ K) excited multiplets\cite{takikawa}.
Figure \ref{fig2_chi} shows the Van-Vleck contribution $\chi_{\textrm{VV}}$ estimated using the above excited multiplet levels for a $J=0$ ground state.
Note that the contribution obtained after subtracting the Van-Vleck contribution from the observed $\chi(T)$ is nearly temperature independent and does not follow the Curie-Weiss form. This strongly suggests that this component comes from the Pauli paramagnetism and indicates the absence of local moment in the metallic phase.
 
At $\Tmn=120$ K the ZFC data exhibit a sharp peak followed by a drop in $\chi(T)$ on cooling. This drop exists for crystals with $\rho(4.2~\mathrm{K})/\rho(300~\mathrm{K}) >500$ and $\rho(4.2~\mathrm{K})/\rho(300~\mathrm{K}) >100$, but not for those with $\rho(4.2~\mathrm{K})/\rho(300~\mathrm{K}) <33$ or the results reported for polycrystalline samples \cite{taira}, wherein $\chi(T)$ returns to the value for the metallic phase. 
Because the size of $\chi(T)$ after the drop for the high 1/RRR samples is close to the one for $\chi_{\rm{VV}}$,
a possible explanation for this drop is the disappearance of the Pauli paramagnetic component as the system becomes insulating by opening a gap at $E_{\rm F}$. 
We note the temperature range in which this drop occurs is concurrent with the initial increase in $\rho$ at $T<\Tmn$. 
The absence of the drop suggests no well defined charge gap, as is consistent with the low value of 1/RRR.

For both FC and ZFC results, the magnetization is almost independent of the magnetic field alignment, indicating that unit cell of the spin structure should be magnetically isotropic. This suggests a non-collinear arrangement of Ir moments, and in particular, together with the commensurate nature found by $\mu$SR measurements \cite{zhao}, the ``all-in all-out'' antiferromagnetic ground state predicted theoretically for the pyrochlore iridates \cite{wan2011, krempa}.

At low temperatures below 5 K, the upturn is observed in $\chi(T)$ for the samples with $\rho(4.2~\mathrm{K})/\rho(300~\mathrm{K}) >500$ and  $\rho(4.2~\mathrm{K})/\rho(300~\mathrm{K}) >100$, while a drop between 3 and 5 K for the one with  $\rho(4.2~\mathrm{K})/\rho(300~\mathrm{K}) <33$. Nearly the same magnitude of the upturn was also observed in the results obtained for our polycrystalline sample and suggests that this comes, not from the surface, but from the magnetic impurity in the bulk induced by off-stoichiometry. 


\begin{figure}[t]
 \begin{center}
  \includegraphics[width=\columnwidth]{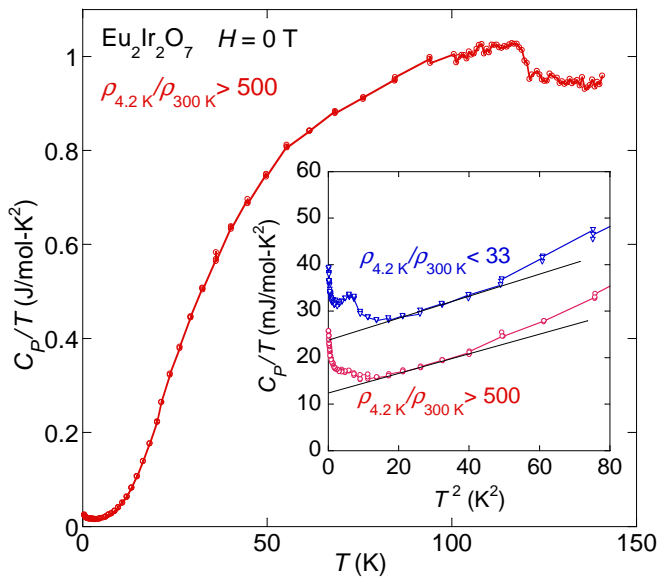}
 \end{center}
 \caption{
(Color online) Temperature dependence of the specific heat divided by temperature, $C_P/T$, of the single crystalline \Eu with 1/RRR $> 500$. Inset: $C_P/T$ vs. $T^2$ for crystals with 1/RRR $> 500$ and with 1/RRR $< 33$. The solid lines are guides to the eye.
}
\label{fig3_c}
\end{figure}

Figure \ref{fig3_c} shows the temperature dependence of the specific heat divided by temperature $C_P/T$ for samples with $\rho(4.2~\mathrm{K})/\rho(300~\mathrm{K}) >500$. 
$C_P/T$ exhibits a clear anomaly at $\Tmn=120$ K providing evidence of the bulk transition. The measurements for samples with  $\rho(4.2~\mathrm{K})/\rho(300~\mathrm{K}) <33$ did not detect any anomaly in $C_P$ and suggests that the off-stoichiometry may broaden the transition, as seen in the $T$ dependence in $\rm{d}\rho/\rm{d}\it T$ (inset of Fig. \ref{fig1_r}).

At low temperatures we estimate the electronic specific heat coefficient, $\gamma$, by plotting $C_P/T$ \emph{vs.} $T^2$. A linear fit is applied between 3 and 6.3 K
yielding $\gamma=12.9(1)$ mJ/mol-K$^2$ and $\gamma=24.9(2)$ mJ/mol-K$^2$ for samples with $\rho(4.2~\mathrm{K})/\rho(300~\mathrm{K}) >500$ and with $\rho(4.2~\mathrm{K})/\rho(300~\mathrm{K}) <33$, respectively. The slope of the fits are the same and gives the Debye $T$ to be 472(2) K. As expected, more conductive samples have larger $\gamma$ due to the presence of additional carriers. The finite $\gamma$ for insulating samples could be associated to the strongly localized states at $E_{\rm F}$ induced by residual disorder. By further assuming a spherical Fermi surface for the localized states to estimate the density of states $N(E_{\rm F})$ at $E_{\rm F}$, the localization length $\xi$ may be calculated\cite{shklovskii, mottdavis, mott}, using the equation $\xi = [21/(\kb T_0 N(E_{\rm F}))]^{1/3}$.
This yields the estimate of $\xi \sim 5$ \AA, which is less than twice of the Ir-Ir ionic distance, consistent with the strong localization.

Below $T=5$ K we find a striking upturn in $C_P/T$ for both insulating and metallic samples. In the same temperature region, the upturn in $\chi(T)$ was observed and therefore, this specific heat anomaly may be attributed to the impurity moments. In particular, for the sample with $\rho(4.2~\mathrm{K})/\rho(300~\mathrm{K}) < 33$, a kink is found at around 3 K in $C_P(T)$, similarly to the one found in $\chi(T)$. This suggests a magnetic phase transition induced by a high concentration of the impurity spins.

Summarizing all the results based on our single crystal study, we find the continuous, most likely second-order, phase transition at $\Tmn = 120$ K from the paramagnetic metal to the antiferromagnetic insulator, which is intrinsic to the stoichiometric \Eu. 
The gap size estimated from the activation law is of the order of 10 meV, which is the same size of the magnetic ordering temperature, $\Tmn$.
This low temperature transport changes dramatically from the insulating behavior to semi-metallic one, strongly depending on the sample stoichiometry. In sharp contrast, the magnetic transition temperature and the temperature dependence of the susceptibility is not affected by the off-stoichiometry. This clearly indicates the strong proximity of the ground state to the metal-insulator transition, and that the magnetic transition is essential for the system to fully open the charge gap when it is stoichiometric. 

According to the recent theoretical calculations for the pyrochlore iridates, various types of interesting ground states have been predicted depending on the strength of the correlation and transfer integrals between the neighboring Ir $5d$ orbitals \cite{balents, ybkim2010, wan2011, kurita, krempa}. Among them, the phases with realistic values of these electronic parameters contain a metal, a topological semimetal, and an antiferromagnetic insulator. Furthermore, both the calculations based on the local spin density approximation (LSDA) as well as the tight binding approximation predicts the magnetic structure to be the ``all-in all-out" type structure reflecting the strong spin-orbit coupling due to Ir $5d$ electrons\cite{wan2011, krempa}. Our observation of magnetic isotropy below $\Tmn$ as well as a commensurate order found by $\mu$SR measurements\cite{zhao} are consistent with the ``all-in all-out" magnetic structure.

Strong excitement made by the theoretical prediction is actually for the possible existence of the topological semimetal phase with Weyl or two-component fermions \cite{wan2011, krempa}. It has been proposed that the topologically protected Dirac points are stabilized, which can be viewed as a three dimensional analog of the electronic structure of graphene, and have an interesting consequence of a metallic surface state.
\cite{wan2011, krempa} For this particular semi-metallic state, the recent detailed calculation indicates that the resistivity should vary as a linear function of $1/T$ and thus $T \rho(T)$ should be constant, reflecting the linearly vanishing density of states nearby the Fermi level.\cite{vishwanath} However, our results of $\rho(T)$ is exponentially divergent on cooling, and excludes the possibility of such an exotic semi-metallic ground state. Instead, all our observations are fully consistent with the prediction for the antiferromagnetic insulator with the ``all-in all-out'' magnetic structure. Furthermore, the theory also suggests its proximity to a magnetic metallic phase with the same ``all-in all-out'' magnetic structure, which is also consistent with our observations.
In addition, the small size of the gap is also the origin why the system is so sensitive to the off-stoichiometry and becomes (semi)metallic by the associated carrier doping.

Finally, our recent study of the resistivity under pressure for \Eu found a quantum phase transition to a diffusive metallic phase from the AF insulator at the ambient pressure.\cite{tafti} According to the theoretical prediction, it is likely that the diffusive metallic phase under pressure may well correspond to the topological semi-metallic phase. Further study on the properties of this state under pressure is highly desired to elucidate the exotic nature of the ground state.

We thank Yo Machida and Yasuo Ohta for the contribution in the early stage of this study. 
We also acknowledge L. Balents, M. Imada, S. Julian, Y. B. Kim, S. Onoda, and M. Takigawa  for useful discussion.
This work is partially supported by Grant-in-Aid
for Scientific Research (No. 21684019) from JSPS, by Grant-in-Aid for
Scientific Research on Priority Areas (No. 19052003) from MEXT, Japan, by Global
COE Programs ``the Physical Sciences Frontier"  MEXT, Japan, and by a Toray Science and Technology Grant. 
E. O.F. acknowledges the Japan Society for the Promotion of Science (JSPS) Fellowship for Foreign Researchers.
The authors also thank the Materials Design and Characterization Laboratory, Institute for Solid State Physics, 
University of Tokyo.


\bibliography{ION}

\end{document}